\documentclass[fleqn,usenatbib]{mnras}

\DeclareRobustCommand{\VAN}[3]{#2}
\let\VANthebibliography\thebibliography
\def\thebibliography{\DeclareRobustCommand{\VAN}[3]{##3}\VANthebibliography}

\usepackage{graphicx}	
\usepackage{amssymb}	

\def\src{IGRJ16327}
\def\srclong{IGR~J16327-4940}

\def\xmm{{\em XMM--Newton}}
\def\inte{{\em INTEGRAL}}

\def\chandra{{\em Chandra}}

\def\gaia{{\em Gaia}}

\def\approxgt{\mathrel{\hbox{\rlap{\lower.55ex \hbox {$\sim$}}
        \kern-.3em \raise.4ex \hbox{$>$}}}}
\def\approxlt{\mathrel{\hbox{\rlap{\lower.55ex \hbox {$\sim$}}
        \kern-.3em \raise.4ex \hbox{$<$}}}}

\def\flux {\mbox{erg cm$^{-2}$ s$^{-1}$}}
\def\lum {\mbox{erg s$^{-1}$}}
\def\nh{$N_{\rm H}$}
\def\ltsima{$\; \buildrel < \over \sim \;$}
\def\lsim{\lower.5ex\hbox{\ltsima}}
\def\gtsima{$\; \buildrel > \over \sim \;$}
\def\gsim{\lower.5ex\hbox{\gtsima}}

\def\ergsec{\hbox{erg s$^{-1}$}}

\def\hcm {\hbox {\ifmmode $ atom cm$^{-2}\else atom cm$^{-2}$\fi}}
\def\arcmin {\hbox{$^\prime$}}
\def\arcsec {\hbox{$^{\prime\prime}$}}

\def \apj {ApJ}
\def \aj {AJ}
\def \apjl {ApJL}
\def \apjs {ApJS}
\def \aap {A\&A}

\def \mnras {MNRAS}
\def \nat {Nature}

\def \nar {New Astronomy Reviews}

\newcommand{\be}{\begin{equation}}

\newcommand{\ee}{\end{equation}}
\newcommand{\ergs}{{~\rm erg\, s^{-1}}}

\newcommand{\swift}{{\emph{Swift}}}

\title[Chandra observation of \srclong]{Probing the nature of the X-ray source IGR J16327-4940 with \chandra}

\author[L. Sidoli et al.]{
L. Sidoli,$^{1}$\thanks{E-mail: lara.sidoli@inaf.it}
V. Sguera,$^{2}$
K. Postnov,$^{3,4}$
P. Esposito,$^{5,1}$
L. Oskinova,$^{6,4}$
and I.A. Mereminskiy$^{7}$
\\
$^{1}$ INAF, Istituto di Astrofisica Spaziale e Fisica Cosmica, via A.\ Corti 12, 20133 Milano, Italy \\
$^{2}$ INAF, Osservatorio di Astrofisica e Scienza dello Spazio, via P.\ Gobetti 101, 40129 Bologna, Italy \\
$^3$ Sternberg Astronomical Institute, Lomonosov Moscow State University, Universitetskij pr. 13, 119234 Moscow, Russia\\
$^4$ Kazan Federal University, Kremlyovskaya 18, 420008 Kazan, Russia \\
$^5$ Scuola Universitaria Superiore IUSS Pavia, Piazza della Vittoria 15, 27100, Pavia, Italy \\
$^6$ Institute for Physics and Astronomy, University Potsdam, 14476 Potsdam, Germany \\
$^7$ Space Research Institute, Russian Academy of Sciences, Profsoyuznaya 84/32, 117997 Moscow, Russia
}

\date{Accepted 2023 September 18. Received 2023 September 15; in original form 2023 July 4}

\pubyear{2023}

\begin{document}
\label{firstpage}
\pagerange{\pageref{firstpage}--\pageref{lastpage}}
\maketitle

\begin{abstract}
  We report on the results of a \chandra\ observation of the source \srclong,
  suggested to be a high mass X-ray binary hosting a luminous blue variable star (LBV). 
  The source field was imaged by ACIS-I in 2023 to search for X-ray emission
  from the LBV star   and eventually confirm this association.
   No X-ray emission is detected from the LBV star, with an upper limit on the X-ray luminosity of 
  L$_{\rm 0.5-10 keV}<2.9(^{+1.6} _{-1.1})\times10^{32}$~\lum\ 
  (at the LBV distance d=12.7$^{+3.2} _{-2.7}$ kpc).
  We detected 21 faint X-ray sources, 8 of which inside the \inte\ error circle.
  The brightest one is the best candidate soft X-ray counterpart of \srclong,  showing a hard power law spectrum  and a flux corrected for the
  absorption UF$_{\rm 0.5-10 keV}$=$2.5\times10^{-13}$~\flux,
  implying a luminosity of $3.0\times10^{33}$  d$_{10~kpc}^2$ \lum.  No optical/near-infrared counterparts have been found.
  Previous X--ray observations of the source field with Swift/XRT and ART-XC did not detect any source consistent with the \inte\ position. 
  These findings  exclude the proposed LBV star as the optical association,
  and  pinpoint the most likely soft X-ray counterpart. 
  In this case, the source properties suggest a low mass X-ray binary, possibly a new member of the very faint X-ray transient class.
\end{abstract}




    \section{Introduction\label{intro}}

The discoveries performed by the \inte\ satellite, surveying the transient hard X-ray sky, have revitalized the field of High Mass X-ray Binaries (HMXBs; \citealt{Kretschmar2019} for a review).  New classes of HMXBs have been unveiled, like the so-called obscured sources \citep{Walter2003} and the Supergiant Fast X-ray Transients (\citealt{Sguera2005, Negueruela2006}).  
The \inte\ catalogs \citep{Bird2010, Bird2016}  still include many unidentified hard X-ray sources 
(detected above 20 keV) whose nature is  unclear and  potentially very intriguing.
 
\srclong\ is one of these sources (\src, hereafter for brevity). It was reported in the \inte/IBIS catalogs \citep{Bird2010, Bird2016} as a faint,
transient hard X-ray source, best detected in the energy band 20--100 keV only once during a short period of enhanced X-ray activity (duration of about 1 day) in March 2005. The source  peak-flux was  2.1 mCrab or 1.6$\times$10$^{-11}$ erg cm$^{-2}$ s$^{-1}$ (20--40 keV ).
Its sky position is at R.A.=248.172$^{\circ}$, dec=$-49.666^{\circ}$ (J2000),
with an error radius of 5\arcmin\ (90\% c.l.).
The source has not been detected in the mosaic significance map obtained by summing up all available \inte\ observations for a total of 4.8 Ms on-source exposure \citep{Bird2016}. 
The inferred 2$\sigma$ upper limit on the persistent emission 
is about 0.2~mCrab or 1.5$\times$10$^{-12}$ erg cm$^{-2}$ s$^{-1}$ (20-40 keV). 
To date, soft X-ray information (below 10 keV) on the source have never been reported in the literature. 

The source was tentatively associated by \citep{Masetti2010} 
with an early-type emission line star (EM$^*$\,VRMF 55; R=15.5 mag), implying its identification with an HMXB.
Further optical investigations revealed the presence of hydrogen and 
iron emission lines typical of luminous blue variables (LBVs; \citealt{Gvaramadze2015}), 
a strong brightness decline (from V$\sim$14.40 mag to V$\sim$15.70 mag in six years) 
and a significant variability in its spectral properties (it became hotter in 2015, compared to 2009).
The star was also found to be surrounded by a circular
nebula (named MN44) in \textit{Spitzer} data \citep{Gvaramadze2010, Gvaramadze2019}, 
another property typical among LBVs, produced by the interaction of the stellar
ejected material with the interstellar medium.
The stellar proper motion of MN44 and its distance
(d=5.2$^{+2.7} _{-1.7}$~kpc, \gaia\ DR2, \citealt{Bailer-Jones2018})  
led \citet{Gvaramadze2018} to conclude that the star is running away from the Galactic plane,
with a trajectory consistent with being born in Westerlund 1, one of the most massive star clusters in our Galaxy. 

We note that an updated distance value is now available and is reported in the \gaia\ EDR3 catalog   
(d=12.7$^{+3.2} _{-2.7}$~kpc, obtained from both the photometry and the parallax; \citealt{Bailer-Jones2021}). 
This new distance rejects the association with Westerlund~1.
We adopt the new \gaia\ EDR3 distance in our paper.

In this paper we mainly report the results of a \chandra\ observation we proposed in order
to unveil the LBV star as the counterpart of \src, possibly detecting X-ray emission from its sky position.

\section{Observation and data reduction}
\label{data}

\chandra\ observed the source sky region on February 11, 2023 (ObsID~26517) with a net exposure time of 9957~s, using ACIS-I in very faint mode. The observation was targeted at the coordinates R.A. (J2000)=248.166417$^{\circ}$, dec (J2000)=$-$49.703797$^{\circ}$ (the sky position of the LBV star) and  aimed at imaging the whole \inte\ error circle.

The data were reduced with the \chandra\ Interactive Analysis of Observation  ({\tt CIAO} 4.14) and CALDB (4.9.8), adopting standard procedures. The tool {\tt chandra\_repro} was applied to reprocess
the level 1 event lists.
Images, exposure and point-spread-function (PSF) maps were produced by the {\tt fluximage} tool in the energy range 0.5-7 keV, 
using the full resolution (pixel size of 0.492\arcsec) and resulting in output images with 2284 by 2275 pixels.

A proper source detection was performed (Sect.~\ref{sect:det}).
For sufficiently bright sources, we performed a spectral and a timing analysis.
The spectroscopy was performed using {\tt xspec} \citep{Arnaud1996}
in  {\tt HEASoft} (\citealt{heasoft}; v.29).
The  model {\tt TBabs} was adopted to account for the low energy absorption in the X-ray spectra. The photoelectric absorption cross sections of \cite{Verner1996} and the interstellar abundances of \cite{Wilms2000} were used. We grouped the spectra adopting 1 count per bin and Cash statistics \citep{Cash1979}.
In the spectroscopy, all uncertainties  are given at 90\% confidence level for one interesting parameter. 
The uncertainty (90\%) on the unabsorbed X-ray fluxes have been calculated using {\tt cflux} in {\tt xspec}.

For fainter sources, we have estimated their fluxes using {\tt srcflux} tool in {\tt CIAO}, with the appropriate response files, assuming an absorbed power law model with a photon index $\Gamma$=2 and an absorption column density N$_{\rm H}$=10$^{22}$~cm$^{-2}$ (results in Table~\ref{tab:sources}). 
 The source-free backgrounds were evaluated locally from annular regions centered on the position of the sources, with an inner and an outer radius of one and five times source extraction radius (90\% PSF radius at 2.4 keV).
When the annular background regions overlap with nearby sources, they have been modified excising the contaminating nearby source region.

The timing analysis has been performed after correcting the arrival times of all events to the Solar System barycenter using {\tt axbary}.

\subsection{Swift}
\label{swiftdata}

We note that the source sky region has never been observed before in soft X-rays, apart from a short \swift/XRT snapshot ($\sim$2.8 ks) performed in June 2012 (ObsID 00032483001), never reported in the literature to date. 
For completeness, we reduced this observation using {\tt xrtpipeline} in {\tt HEASoft} and extracted an image and its exposure map in the total band (0.3-10 keV). We did not detect any source within the XRT field-of-view (FOV). Using {\tt sosta}  in {\tt ximage} we derived a 3$\sigma$ upper limit of 4.2$\times$10$^{-3}$~s$^{-1}$ (0.3-10 keV) at the LBV star position.
Using the  appropriate calibration files for the date of this observation, this count rate translates into an upper limit on the unabsorbed flux of $\sim$5$\times$10$^{-13}$~\flux\ (0.5-10 keV), assuming a power law  model with a photon index of 2 and an absorbing column density \nh=10$^{22}$~cm$^{-2}$.

\subsection{ART-XC}
\label{artxc}

In the 4-12 keV energy range, no new sources in the \inte\ error region were reported in the first year All-sky ART-XC catalog \citep{2022A&A...661A..38P} performed in 2019-2020 by Mikhail Pavlinsky ART-XC telescope onboard the Spectrum-Roentgen-Gamma mission \citep{2021A&A...650A..42P}. In 2022-2023, the Norma Arm region have been visited for a total time of about two ks during the new Galactic plane ART-XC X-ray survey. The upper limit 4-12 keV from possible X-ray sources in the \inte\ error radius is $F_{4-12\,\mathrm{keV}}<4\times 10^{-13}$ \flux (90\% c.l.) assuming a Crab-like spectrum. This upper limit yields an X-ray luminosity limit of L$_{4-12\,\mathrm{keV}}<5\times 10^{33}(d^2_{10\,kpc})$ \ergsec, an order of magnitude looser than the \chandra\  limit (see below) but taken at different times than our \chandra\ observations.

\section{Results}
\label{sect:res}

In the following we report on the \chandra\ results, while the archival \swift\ observation and ART-XC limits are not discussed further.

\subsection{Source detection}
\label{sect:det}

\begin{figure*}
\begin{center}
\includegraphics[width=10.cm,angle=0]{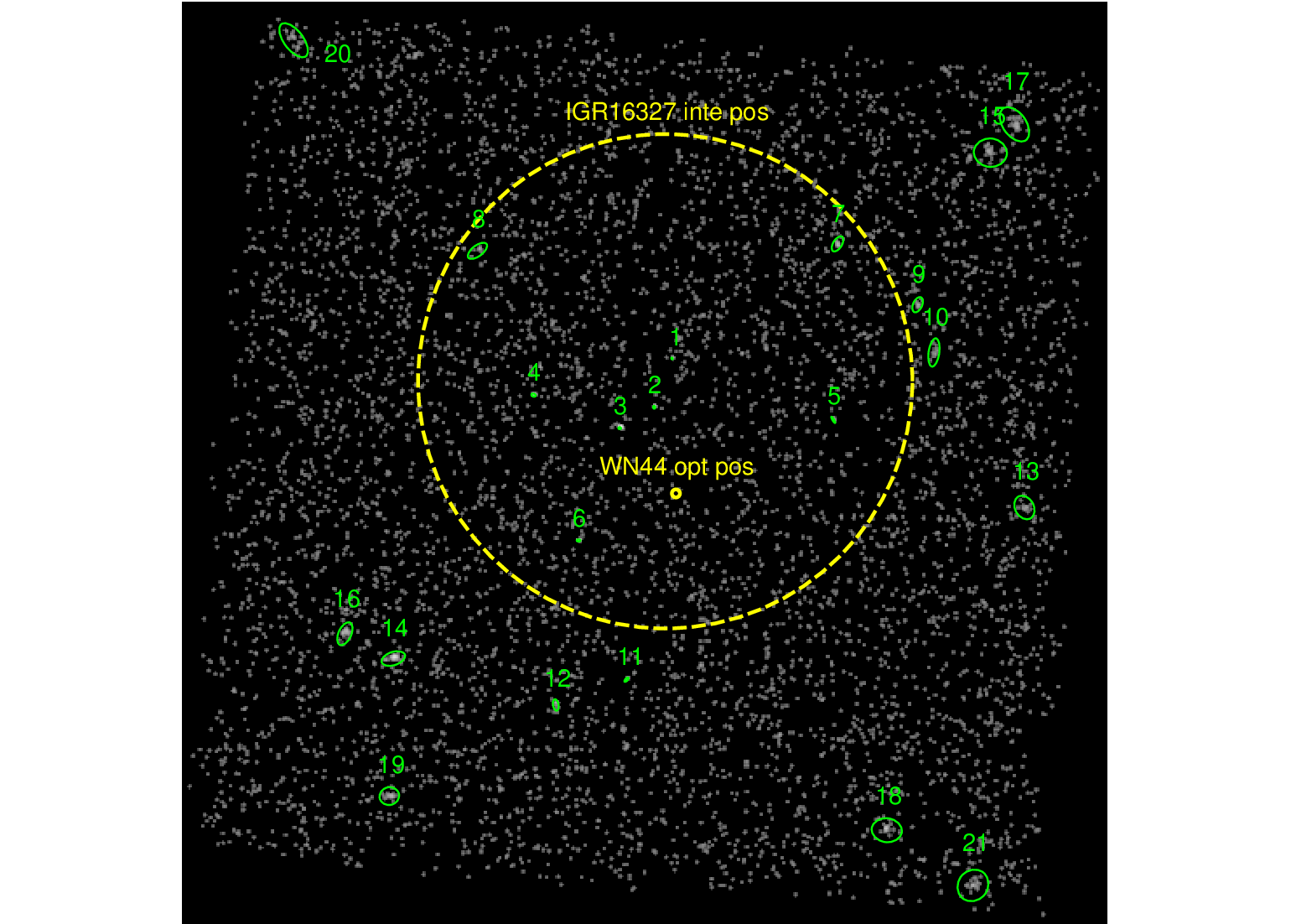}
   \includegraphics[width=7.5cm,angle=0]{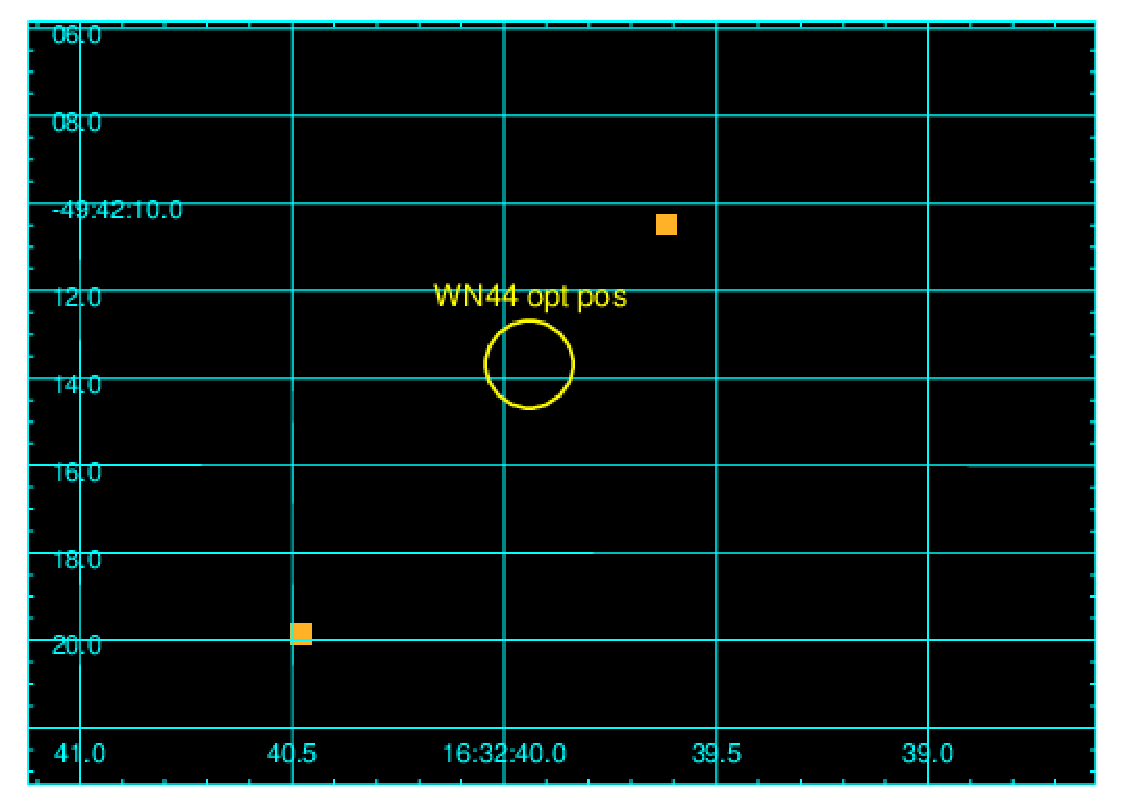}
  \caption{{\em Left panel}: \chandra\ ACIS-I observation (0.5-7 keV) of the \src\ field, smoothed with a Gaussian function, for presentation purposes only.
  The dashed yellow large circle marks the \inte\ source error circle (radius R=5\arcmin), while the
  small circle indicates the position of the LBV star (WN44). 
  The green ellipses mark the 21 \chandra\ sources detected in the energy band 0.5-7 keV. 
  The brightest source inside the \src\ error circle is n.~3. 
  {\em Right panel}: a close-up view of the on-axis position of the ACIS-I observation is shown, together with the optical position of LBV star marked by the yellow circle (R=1\arcsec). Equatorial coordinates (J2000) are reported. No \chandra\ detection is consistent with the LBV star position.
}
\label{fig:acis}
\end{center}
\end{figure*}

We performed a source detection by means of the {\tt wavdetect} algorithm in {\tt CIAO}, which uses
``Mexican hat'' wavelet functions with scales of 1, 2, 4, 6, 8, 12, 16, 24 and 32. 
We adopted a detection threshold equal to one over the area of the image in pixels, translating into approximately
one spurious source (e.g., \citealt{Nandra2005}) inside the FOV. 
This resulted into 21 sources detected in the broad energy band 0.5-7 keV (Fig.~\ref{fig:acis}; Table~\ref{tab:sources}).
Eight sources are located within the \src\ error circle, but none of them is consistent with the sky position of the LBV star (Fig.~\ref{fig:acis}, right panel).
We note that, if we perform a source detection with {\tt wavdetect} in the energy bands 0.5-1.2 keV (soft), 1.2-2.0 keV (medium) and 2.0-7.0 keV (hard), usually assumed in the \chandra\ Source Catalog \citep{Evans2010},  none is detected in the soft band and most of them are detected in the hard band only.

We explored the possibility to correct the absolute astrometry using the tools {\tt wcs\_match} and {\tt wcs\_update}.
They compute the fine astrometic translation shift between the list of \chandra\ detections and source lists at longer wavelenghts.
We have chosen the \gaia\ DR2 catalog of optical sources which can be retrieved online by means of 
the software {\tt SAOImage} (ds9; \citealt{Saoimage}) included in {\tt CIAO}. 
Using {\tt wcs\_match} (with a match radius of 2\arcsec) to cross-correlate the \chandra\ detections obtained with {\tt wavdetect},
we got six matches with the  \gaia\ catalog.
However, since only two of them are located within 3\arcmin of the aimpoint (sources n.\,1 and 6),
we decided to avoid making any astrometric correction. 
We would like to note that there are not known secure associations of X-ray sources with other catalogs, as this region of the sky
has never been observed at soft X-rays by other space missions, except for the short \swift\ observation (Sect.~\ref{swiftdata}) which did not detect any source.
Therefore, we assume here an absolute positional accuracy of 0.92\arcsec\ (90\% limit, ACIS-I).\footnote{https://cxc.harvard.edu/cal/ASPECT/celmon} 
We note that the uncertainties on the sky coordinates reported in Table~\ref{tab:sources} 
are the statistical ones resulting from {\tt wavdetect}.
In the same table we report also the angular distance from the centroid of the \inte\ error circle ($\theta_{igr}$) 
and the unabsorbed fluxes (0.5-7 keV) calculated using the tool 
{\tt srcflux} in {\tt CIAO}, adopting a power law model with a photon index of 2 and an absorbing column density of 10$^{22}$~cm$^{-2}$.

\begin{table*}  
  \caption{List of sources detected with {\tt wavdetect} (0.5-7 keV). Fluxes UF$_{0.5-7~\rm keV}$ are corrected for the absorption and calculated with {\tt srcflux} (uncertainties on fluxes are at 90\% c.l.).    $\theta_{igr}$ is the angular distance from the \inte\ centroid of \src. ``P" is the probability (\%) of spurious association of the \chandra\ sources with \src.
}
\label{tab:sources}
\vspace{0.0 cm}
\begin{center}
\begin{tabular}{ccclllllr} \hline
\noalign {\smallskip}
Source n. &   R.A. (J2000)      &  dec   (J2000)      &     error on R.A. &  error on dec &      Net counts   &  $\theta_{igr}$ &  UF$_{0.5-7~\rm keV}$  &   P \\
        &      (deg)          &   (deg)             &          (arcsec) &    (arcsec)   &       (0.5-7 keV)   &  (\arcmin)  &   (10$^{-13}$ \flux)   &  (\%)  \\
\hline
 1  &         248.16844   &         -49.658329  &       0.18        &       0.15    &        5.80             &  0.48   & 0.21$^{+0.17} _{-0.11}$  &   100 \\
 2  &         248.17773   &         -49.674763  &       0.20        &       0.17    &        11.0             &  0.57   &  0.51$^{+0.34} _{-0.24}$ &  97  \\
 3  &         248.19571   &         -49.681797  &       0.12        &       0.07    &        46.9             &  1.32   &  1.7$\pm{0.4}$  & 53  \\
 4  &         248.24059   &         -49.670654  &       0.41        &       0.19    &        10.7             &  2.68   & 0.36$^{+0.22} _{-0.16}$   &  99 \\
 5  &         248.08443   &         -49.678978  &       0.54        &       0.56    &        4.70             &  3.49   & 0.18$^{+0.17} _{-0.10}$  &  100  \\
 6  &         248.21715   &         -49.719669  &       0.27        &       0.10    &        17.9             &  3.67   & 0.48$^{+0.25} _{-0.19}$  &  97  \\
 7  &         248.08231   &         -49.619652  &       1.02        &       0.85    &        10.4             &  4.46   &    0.43$^{+0.27} _{-0.19}$  &  98  \\
 8  &         248.26971   &         -49.621994  &       2.01        &       1.09    &        7.10             &  4.62   &    0.24$^{+0.22} _{-0.14}$  &  100 \\
 9  &         248.04057   &         -49.640110  &       1.05        &       0.97    &        8.10             &  5.34   &   0.32$^{+0.24} _{-0.17}$  &  100 \\
10  &         248.03203   &         -49.656200  &       1.06        &       1.72    &        8.50             &  5.47   &  0.30$^{+0.24} _{-0.16}$ &  100  \\
11  &         248.19220   &         -49.766590  &       0.53        &       0.36    &        5.70             &  6.09   &  0.32$^{+0.27} _{-0.18}$  &   100  \\
12  &         248.22911   &         -49.775108  &       0.64        &       0.81    &        7.00             &  6.91   &   0.24$^{+0.22} _{-0.14}$  &  100 \\
13  &         247.98470   &         -49.708397  &       1.59        &       1.20    &        11.9             &  7.70   &  0.48$^{+0.33} _{-0.24}$  &  100 \\
14  &         248.31387   &         -49.759392  &       0.95        &       0.38    &        54.5             &  7.85   &  2.1$\pm{0.5}$  & 76  \\  
15  &         248.00288   &         -49.588818  &       1.98        &       1.11    &        17.4             &  8.04   &  0.88$^{+0.42} _{-0.34}$  &   99 \\
16  &         248.33914   &         -49.750984  &       0.93        &       0.89    &        22.5             &  8.25   &   1.00$^{+0.43} _{-0.30}$  &  98 \\
17  &         247.99036   &         -49.579239  &       1.71        &       1.30    &        20.1             &  8.77   &   1.16$^{+0.49} _{-0.39}$   &  98 \\
18  &         248.05620   &         -49.817280  &       1.89        &       0.97    &        20.7             &  10.1   & 0.90$^{+0.40} _{-0.30}$  &  100 \\
19  &         248.31612   &         -49.805748  &       1.28        &       0.75    &        18.6             &  10.1   &   0.96$^{+0.43} _{-0.34}$   & 100 \\
20  &         248.36510   &         -49.550873  &       1.96        &       1.50    &        24.8             &  10.2   &   1.22$^{+0.56} _{-0.44}$ &  99  \\
21  &         248.01111   &         -49.835835  &       1.91        &       1.26    &        18.4             &  11.9   &   1.10$^{+0.46} _{-0.40}$  &  100  \\
\hline
\end{tabular}
\end{center}
\end{table*}

Since none of the detected \chandra\ sources is consistent with the position of the LBV star, we have estimated an
upper limit  on the ACIS-I net count rate (99\% c.l.; \citealt{Kraft1991}) of R$<$$4.6\times10^{-4}$~count~s$^{-1}$ (0.5-7 keV).
Assuming an absorbed power law model with a photon index $\Gamma$=2 and an absorption column density
N$_{\rm H}$=10$^{22}$~cm$^{-2}$ (appropriate for the interstellar extinction to the star), we obtain 
an upper limit on the flux corrected for the absorption 
UF$_{0.5-10~\rm keV}$$<$1.5$\times10^{-14}$~\flux.
Assuming the LBV distance d=12.7$^{+3.2} _{-2.7}$ kpc, it translates into an upper limit
on the X-ray luminosity of L$_{0.5-10~\rm keV}$$<2.9(^{+1.6} _{-1.1})\times10^{32}$~\lum\ (99\% c.l.).
We note that these uncertainties are calculated adopting the maximum and minimum values of the LBV star distance.

\subsection{Spectroscopy}
 
\begin{figure*}
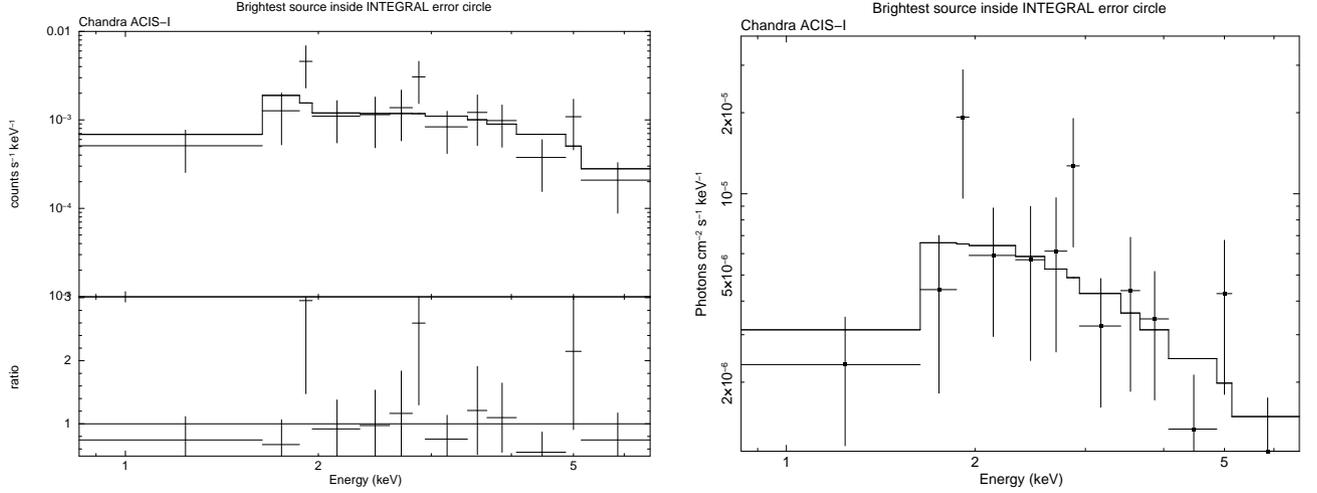

\begin{center}
\includegraphics[width=6.5cm,angle=-90]{lda_rat_src1_pow.ps} 
\vspace{1.0cm}
\includegraphics[width=6.5cm,angle=-90]{ufs_src1_pow.ps}  
\caption{Spectrum of source n.\,3, fitted with an absorbed power law model.
  The left panel shows the count spectrum together with the residuals. 
  The photon spectrum is shown on the right.
  The spectra have been re-binned for presentation purposes only.
}
\label{fig:spec1}
\end{center}
\end{figure*}

We extracted a spectrum from the brightest X-ray source detected within the \inte\ error radius of \src, the source n.\,3 in Table~\ref{tab:sources}.
A simple absorbed power law is already a good deconvolution of the spectrum. 
We report the spectral parameters in Table~\ref{tab:spec1}, where we have calculated the X-ray luminosity using a distance value in units of 10 kpc, for a rapid re-scaling, since the distance is unknown. In fact, this source has no optical/NIR counterparts (see Sect.~\ref{sect:other}). The spectrum is shown in Fig.~\ref{fig:spec1}.
   
We extracted the spectrum also from other \chandra\ sources for which a meaningful spectroscopy can be performed.
We report in Table~\ref{tab:otherspec} the results.

\begin{table}
  \caption{Spectroscopy of the brightest X-ray source within the \inte\ error circle (source n.\,3) adopting an absorbed power law model.
}
\label{tab:spec1}
\vspace{0.0 cm}
\begin{center}
\begin{tabular}{ll}  
 \hline
\noalign {\smallskip}
Param.                          &                          \\
\hline
 N$_{\rm H}$                    &  $2.4^{+4.2} _{-2.3}$  $\times10^{22}$ cm$^{-2}$   \\
Photon index $\Gamma$           & $2.0^{+1.8} _{-1.3}$   \\
UF$_{0.5-10~\rm keV}$           &  $2.5\times10^{-13}$ \flux  \\
 C-Stat                         &     35.74 (36 dof)                   \\
L$_{0.5-10~\rm keV}$  &  $3.0\times10^{33}$  d$_{10~kpc}^2$ \lum     \\
\hline
\end{tabular}
\end{center}
\end{table}

\begin{table*}
  \caption{Spectroscopy of other \chandra\ sources, with net counts in excess of 20 (0.5-7 keV; Table \ref{tab:sources}), assuming an absorbed power law model. 
}
\label{tab:otherspec}
\vspace{0.0 cm}
\begin{center}
\begin{tabular}{rlllll} \hline
\noalign {\smallskip}
Source n.               &  N$_{\rm H}$            &  $\Gamma$               &   UF$_{0.5-10~\rm keV}$     &    L$_{0.5-10~\rm keV}$    &     C-Stat    \\
                        &  (10$^{22}$ cm$^{-2}$)  &                  &    (\flux)                   &         (\lum)             &               \\
\hline
14                      &  $4.1^{+5.2} _{-4.1}$   &  $2.0^{+1.7} _{-1.5}$  &  $3.8^{+8.0} _{-2.8}\times10^{-13}$  &     4.5$\times10^{33}$ d$_{10~kpc}^2$    &   40.84 (41 dof)  \\
16    &       $<$1.7            &  $1.8^{+1.3} _{-0.6}$  &  $7.5^{+10.5} _{-2.5}\times10^{-14}$  &    9.0$\times10^{32}$ d$_{10~kpc}^2$           &   21.44 (19 dof)   \\
17    &         1.0 (fixed)        &  $0.7^{+1.0} _{-1.0}$  &  $1.5^{+1.1} _{-0.6}\times10^{-13}$  &    1.8$\times10^{33}$ d$_{10~kpc}^2$           &   13.70 (22 dof) \\
18    &       1.0 (fixed)           &  $0.9^{+1.2} _{-1.2}$  &  $1.3^{+1.2} _{-0.5}\times10^{-13}$  &    1.6$\times10^{33}$ d$_{10~kpc}^2$           &   11.11 (16 dof) \\
20    &         1.0 (fixed)        &  $1.9^{+1.1} _{-1.0}$  &  $1.2^{+0.8} _{-0.4}\times10^{-13}$  &    1.4$\times10^{33}$ d$_{10~kpc}^2$           &   16.53 (18 dof) \\
\hline
\end{tabular}
\end{center}
\end{table*}

\subsection{Temporal analysis}

We examined the light curve (0.5--7 keV) of source n.\,3 trying several statistical tests on binned and unbinned data, but we did not found evidence for flaring activity or any significant variability in general (e.g., $\chi$--square or Kolmogorov--Smirnov probability of constancy tests never resulted in variability larger than 2$\sigma$). The timing analysis of the barycentered events files did not yield any significant signal as well, with 3$\sigma$ limits on the pulsed fraction of $\sim$100\% for a sinusoidal periodic modulation between $\sim$6 and 5000\,s (which is not surprising, giving the very few counts). We performed the same checks on the data of source n.\,14, the only one with a comparable number of net photons, but again with no positive results or meaningful upper limits.

\section{Discussion}
\label{sect:discussion}

Here we discuss the nature of the source \src\ in light of our \chandra\ results.
First of all, we did not detect any \chandra\ source at the position of the LBV star.
These findings lead to two possibilities: either the LBV star is the true optical counterpart
of \src\ but displays a very low X-ray luminosity state, or one of the \chandra\ sources
detected inside the \inte\ error circle is the correct association. 
We will discuss these two cases in Sect.~\ref{sect:lbv}  and  Sect.~\ref{sect:other}, respectively.

\subsection{Is the LBV star still a viable optical counterpart?}
\label{sect:lbv}

LBV stars are known to be at the transitional stage of evolution of very massive OB stars toward Wolf-Rayet (WR) stars characterized by a huge mass loss of about $10^{-4} M_\odot$ per year \citep{1994PASP..106.1025H}. X-ray emission from stellar winds were detected by \chandra\ and \xmm\ from only a few single LBV stars \citep{2012A&A...538A..47N}. The \chandra\ upper limit from WN44 obtained above is consistent with those for other LBVs. 
Therefore, the association of WN44 
as a single LBV with the \inte\ source cannot be reliably established.

In principle, X-ray emission from an LBV can be due to accretion of stellar wind onto the secondary compact component in a binary system. From evolutionary considerations, the secondary component in this case must be the remnant of a more massive primary component, i.e. most likely a black hole.

If the hard \inte\ flare from \src\ were associated with the WN44, the maximum hard X-ray luminosity would be L$_{20-40\,\mathrm{keV}}\approx 2\times 10^{35}(d^2_{10\,kpc})$ \ergsec. While not excluded in principle, such flares have not been detected from other single LBVs. If the flare were due to non-stationary accretion onto the possible compact binary counterpart, the peak X-ray luminosity would correspond to a maximum accretion rate of $\dot M_a^{peak}\sim 2\times 10^{15}\eta_{0.1}$~[g~s$^{-1}$] $\sim (3\times 10^{-11})\eta_{0.1}^{-1} [M_\odot$~yr$^{-1}$] assuming the standard 10\% accretion luminosity $L=\eta_{0.1}\dot M c^2$. This is much smaller than the expected Eddington-limited mass accretion rate in X-ray outbursts.

Very small upper limits on the quiescent \chandra\ X-ray luminosity from the \inte\ error circle, L$< 3\times 10^{32} \ergs$, suggest at least a three order of magnitude lower mass accretion rate onto a compact object than
the \inte\ flare discussed above,
$\dot M_a^{qsc}\sim 2\times 10^{12}\eta_{0.1}$~[g~s$^{-1}$] $\sim (3\times 10^{-14})\eta_{0.1}^{-1} [M_\odot$~yr$^{-1}$].
The Bondi-Hoyle-Littleton accretion rate from stellar wind  from an LBV star with spherical mass-loss rate $\dot M_{LBV}$ and the wind velocity $v_w$
onto a compact object with mass $M_x$ and Bondi radius $R_B=2GM_x/v_w^2$ in a circular orbit with radius $a$, is 
$\dot M_{BHL}\approx  (1/4) \dot M_{LBV}(R_B/a)^2$. This is an upper limit of the actual accretion rate in the case of quasi-spherical hot radiation-inefficient accretion flows, with the reduction factor being $b=\dot M_a^{qsc}/\dot M_{BHL}\sim (R_0/R_B)^{0.7}$ \citep{2023arXiv230509737X}. Assuming the compact object to be a 10 $M_\odot$ black hole, the accretor radius is $R_0\approx 3 \times 10^6$~cm and the characteristic Bondi radius is $R_B=3\times 10^{12}[\mathrm{cm}] (M_x/10M_\odot)/(v_w/300[\mathrm{km\,s^{-1}]})^2$. Therefore, for a 10 $M_\odot$ black hole the reduction factor in a hot accretion flow can be $b\sim 10^{-4.2}$. Thus, for the standard 10\% radiation efficiency the bolometric accretion luminosity L$_{bol}=\eta_{0.1}b\dot M_{BHL}c^2$.

Taking the \chandra\ upper limits to the accretion luminosity, this would suggest a Bondi radius to orbital separation ratio of 
$(R_B/a)^2> 10^{-8}/(b\eta_{0.1}\dot M_{-5})$ for the fiducial $\dot M_{LBV}= \dot M_{-5} 10^{-5}[M_\odot $~yr$^{-1}$].  Therefore, the possible orbital separation in such a binary should be  $a> 10^4 R_B (b\eta_{0.1}\dot M_{-5})^{1/2}\sim 100 R_B$. Unfortunately, the scarcity of information about the source does not allow us to further constrain the possible binary parameters. Nevertheless, even in the case of 
inefficient hot accretion in a binary with a black hole component, which cannot be excluded by our \chandra\ and ART-XC X-ray limits, the appearance of a sub-luminous outburst detected by \inte\ seems unnatural for non-stationary accretion as in X-ray novae.

\subsection{Alternative counterparts}
\label{sect:other}

 The lack of detection of soft X--rays at the position of the LBV star 
 suggests alternative possibilities to uncover the nature of the \inte\ source. 
 The best candidate soft X-ray counterpart of \src\ is the brightest \chandra\ source detected within the \inte\ error circle: the source n.\,3 in Table~\ref{tab:sources}.
 We calculated the probability of association by chance between this source and the \inte\ source as in the following. 
In fact, assuming the number-flux distribution ($\log{N}$-$\log{S}$) obtained for sources in the Norma Arm region, including the AGNs \citep{Fornasini2014, Tomsick2020}:  

\begin{equation}
N(>UF_{2-10~\rm keV}) = 36(UF_{2-10~\rm keV}/10^{-13})^{-1.24}\,{\rm deg}^{-2}
\end{equation}
where UF$_{2-10~\rm keV}$ is the flux in the energy range 2--10 keV,  corrected for the absorption, we find  N=0.55 sources brighter than source n.~{\bf 3} (UF$_{2-10~\rm keV}$=1.3$\times10^{-13}$~\flux; Table~\ref{tab:spec1}) expected by chance within the \inte\ error circle (5\arcmin\ radius). 
The probability that the \chandra\ source is a chance coincidence association is: 

\begin{equation}
P = 1 - e^{-N(>UF_{2-10~\rm keV})~\pi~\theta^{2}}
\end{equation}
where $\theta$ is the angular distance (in units of degrees) of the \chandra\ source from the \inte\ position ($\theta$=5\arcmin\ for sources within the \inte\ error circle, while $\theta$=$\theta_{igr}$, as listed in Table~\ref{tab:sources}, for sources at larger distances).
This implies a probability of $\sim$40\% of spurious association of \chandra\ source n.~{\bf 3} with \src. In Table~\ref{tab:sources} we list the
probability of spurious association with \src\ for all \chandra\ sources (P$\sim$53\% for source n.{\bf 3} if we adopt the flux obtained with {\tt srcflux}).
For all other sources this probability is much larger (and in excess of 99\% for most of them).

Given these results, the source n.\,3  is the best soft X--ray counterpart of \src, among the detected \chandra\ sources. 
Our spectral analysis indicates a quite hard X-ray emission,  consistent with an X-ray binary in quiescence 
(Table~\ref{tab:spec1}), thus supporting the association with \src. 

For what concerns longer wavelengths, we searched all 
VizieR catalogues (via HEASARC  and the ESO portal), resulting in no association of source n.\,3 with any counterpart.
Searching the VVV survey data we obtained the following limiting magnitudes: Y=20.2, J=20.1, H=19.2 and Ks=19.0 ($\pm{0.1}$ mag).
However, a lower limit to the X-ray-to-optical ($V$ magnitude) flux ratio (F$_{\rm X}/F_{\rm opt}$) can help in constraining its nature. 
We estimate this ratio adopting the limiting magnitude of the USNO-B catalogue \citep{Monet2003} searched for counterparts (V\gtsima21~mag)
and the observed X-ray source flux 
F$_{\rm X}$$=4.6\times10^{-14}$~\flux\ measured from the ACIS-I spectrum in the 0.3-3.5 energy band \citep{Maccacaro1988}, 
as follows:   
$\log (F_{\rm X}/F_{\rm opt}) = \log (F_{\rm X}) + V/2.5 + 5.37$, obtaining $\log (F_{\rm X}/F_{\rm opt})$\gtsima0.433.
This value rules out a stellar coronal origin \citep{Maccacaro1988} as well as a cataclysmic variable \citep{Kuulkers2006} and a HMXB \citep{Tauris2006}, leaving open the possibility of a low mass X-ray binary (LMXB) or an active galactic nucleus (AGN; \citealt{Maccacaro1988}).

If we assume that this \chandra\ source is the true soft X-ray counterpart of \src, it shows an amplitude of variability of about a factor of 270 between the \inte\ luminosity re-scaled to the 0.5-10 keV energy band (L$_{0.5-10~\rm keV}$ $\sim$ $8\times10^{35}$  d$_{10~kpc}^2$ \lum) and the \chandra\ value ($3\times10^{33}$  d$_{10~kpc}^2$ \lum; Table\,\ref{tab:spec1}), assuming the same Crab-like power law spectrum.
The low X-ray luminosity in outburst and the observed dynamic range is reminiscent of the so-called very faint X-ray transients (VFXTs; \citealt{Wijnands2006, Degenaar2010, Wijnands2015}).
This is a sub-class of LMXBs showing low outburst luminosities (L$_X$=10$^{34}$-10$^{36}$~\lum) and quiescent X-ray emission at a level of L$_X$=10$^{30}$-10$^{33}$~\lum. Some VFXTs certainly harbour a neutron star, given the detection of type I bursts (e.g., XMMU~J174716.1-281048, \citealt{DelSanto2007}). The duration of their outbursts can be very variable among the members of this class, from several days to a few years. The mechanism at work producing their under-luminous X-ray outbursts is still under debate.

About the possibility of an AGN origin, the absorbing column density measured from the \chandra\ spectrum is consistent with the total Galactic value towards the source. Therefore, the intrinsic absorption of a putative AGN must be low, indicative of either an un-obscured (type 1) AGN or an intermediate type 2 Seyfert (1.8-1.9; e.g. \citealt{Salvati2000, Alexander2016}).
Although the power law photon index of the \chandra\ spectrum is consistent with an AGN origin, the amplitude of the long-term X-ray flux variability is more difficult to reconcile with this possibility. 
  X-ray outbursts with a similar dynamic range  have been observed in the so-called quasi-periodic erupting AGN sources, possibly produced by the tidal disruption of a stellar-mass object by the central supermassive black hole \citep{Miniutti2019}. However, all the known sources displayed ultra-soft X-ray emission.

The lack of any counterpart at longer wavelengths prevents us from reaching a conclusive answer.
However, while it is true that \inte\ observations have disclosed many AGNs through the Galactic plane (e.g. \citealt{Malizia2020}) we note that the low Galactic latitude (b=$-$1.16$^{\circ}$) favours a Galactic object.

 \section{Conclusions}
 \label{sect:concl}

\src\ was suggested to be a candidate HMXB hosting an LBV donor star. 
We proposed and performed a \chandra\ observation of the source sky position to test this hypothesis,
leading to no X-ray detection at the LBV sky position. 
The measured upper limit to its X-ray flux of 
 UF$_{\rm 0.5-10 keV}$$<$1.5$\times$10$^{-14}$~\flux\ (99\% confidence level) allowed us to rule out the association of the \inte\ source with the LBV star.
 
 Eight faint \chandra\ sources were detected inside the \inte\ error circle.
 Among them, we propose that the brightest one is the most likely soft X-ray counterpart of \src. 
 It displays a hard X-ray spectrum, consistent with an X-ray binary in quiescence, and no optical or NIR associations. Its X-ray luminosity is  $3.0\times10^{33}$~d$_{10~kpc}^2$~\lum. 
 If this is the true counterpart of \src, it implies a flux dynamic range of a factor of at least 270, compared with the X-ray emission of the \inte\ source extrapolated to the 0.5-10 keV energy range, adopting a Crab-like spectrum.
 These properties constrain the  nature of \src\ either as an AGN or, more likely - given the X-ray flux variability and the location on the Galactic plane - an LMXB. In particular, we propose the identification of \src\ with a new member of the class of the VFXTs.
 
In any case, whatever the true counterpart, thanks to the \chandra\ observation we have obtained a clear evidence of the transient (or highly variable) nature of the source \src.


\section*{Acknowledgements}   
This work is based on observations performed with the \chandra\ satellite. 
We made use of the software provided by the \chandra\ X-ray Center (CXC) in the application package {\tt CIAO}, including {\tt SAOImage}.
We made use of the High Energy Astrophysics Science Archive Research Center (HEASARC), 
a service of the Astrophysics Science Division at NASA/GSFC.
This research has made use of the VizieR catalogue service, by means of HEASARC,  CDS and ESO portal. 
This work has made use of data from the European Space Agency (ESA) mission Gaia (https://www.cosmos.esa.int/gaia),
processed by the Gaia Data Processing and Analysis Consortium (DPAC, https://www.cosmos.esa.int/web/gaia/dpac/consortium).
Funding for the DPAC has been provided by national institutions, in particular 
the institutions participating in the Gaia Multilateral Agreement.
LS thanks Manuela Molina for interesting discussions. The authors thank the anonymous referee for the very constructive report.

\section*{Data Availability}

The \chandra\ data analysed here (ObsID~26517) are
publicly available by means of the CXC or the HEASARC.



\bibliographystyle{mnras}








\bsp	
\label{lastpage}
\end{document}